\documentclass[twocolumn,prb,amsmath,amssymb,amsfonts,superscriptaddress,floatfix,showpacs]{revtex4}

\usepackage{graphicx}
\usepackage{dcolumn}
\usepackage{bm}
\usepackage{rotating} 
\usepackage{color}

\begin{document}

\title{ Self-interaction Corrected Local Spin Density Theory of
      5f Electron Localization in Actinides }

\author{A. Svane}
\affiliation{Department of Physics and Astronomy, University of Aarhus, DK-8000 Aarhus C, Denmark}
\author{L. Petit}
\affiliation{Computer Science and Mathematics Division, and Center for Nanophase Materials
 Sciences, Oak Ridge National Laboratory, Oak Ridge, Tennessee 37831, USA}
\author{Z. Szotek}
\affiliation{Daresbury Laboratory, Daresbury, Warrington WA4 4AD, United Kingdom}
\author{W. M. Temmerman}
\affiliation{Daresbury Laboratory, Daresbury, Warrington WA4 4AD, United Kingdom}

\date{\today}

\begin{abstract}
The electronic structures of the actinide elements U, Np, Pu, Am, Cm and Bk 
are investigated within the self-interaction corrected local spin density
approximation. This method allows to describe a dual character of the $5f$ electrons,
some of which occupy localized and core-like
states, while the remaining $5f$ electrons hybridize and form 
bands. Based on energetics the calculations predict delocalization/paramagnetism in the early
actinides, and localization/anti-ferromagnetism in the later actinides. The corresponding calculated equilibrium
volumes are in agreement with the experimental values.
For Pu and Am, the method wrongly predicts magnetic ordering, but we find that
the paramagnetic state gives a better description of cohesive properties. 
Under compression, in the later actinides, a localization-delocalization 
transition happens gradually as more and more $f$ electrons become band-like with
decreasing volume. Pu is already at this transition point at ambient conditions.
Delocalization sets in for Am and Bk at a compression of $V\sim 0.75V_0$, 
for Cm at $V\sim 0.60V_0$,
where $V_0$ is the equilibrium volume, and the transition is complete 
for $V\sim 0.4-0.5 V_0$ in these three elements. 
\end{abstract}

\maketitle

\section{Introduction}

The electronic structure of actinides remains an active area of research both 
experimentally\cite{freeman,freelan,borje-ac,borje,hecker,benedict2,pu-pes,havela,durakiewicz,lashley,landerCm,tobin}
 and theoretically.\cite{HLS,brooks,olleE,soderlind1,penicaud,
soderlind,singh,olle-delta,savrasov,savrasov1,bouchet,SSC,dai,Niklasson,soderlind2,shorikov,shick,savrasov2,dmft} 
The intricate nature of the $f$-electrons, their
sensitivity to external probes, like temperature and pressure, and their magnetic ordering are
at the center of interest. Already in the seventies it emerged that with respect to volume behaviour 
one had to distinguish between the early actinide elements, with their bonding properties resembling those of
the transition metals, and the late actinides, where the lattice parameters remain relatively unchanged,
reminiscent of the localized $f$-states of the rare earth series.~\cite{borje}
Pu is on the border line between these two opposing pictures with a particularly rich phase
diagram.\cite{hecker} 
These trends observed across the actinide series can be correlated with an increase of the on-site 
Coulomb correlation energy of the $f$-electrons which gradually conquers the band formation energy 
as one progresses through the series. 
With applied external pressure the late actinides exhibit
volume collapses associated with
structural phase transitions signaling the onset of $f$-bonding.\cite{benedict2}

Most theoretical studies on the actinide elements\cite{HLS,brooks,olleE,soderlind1,penicaud,soderlind,singh,olle-delta,
savrasov,savrasov1,bouchet,SSC,dai,Niklasson,soderlind2,shorikov,shick,savrasov2,dmft,shickagain}
are based on density functional theory (DFT), in either its local density or generalized gradient approximation,
and their overall conclusion is that the early actinides, Th, Pa, U, and Np, are well described by the conventional band
theory, as their $f$-electrons contribute to bonding, and the equilibrium structures and lattice parameters are 
in good agreement with experiments.\cite{singh}
However, for the later actinides the band picture fails, and additional assumptions or parameters derived
from experiments need to be invoked, diminishing
the predictive power of the approach.
Thus, from Cm to Es, the lattice parameters are found to be in better agreement 
with experiment when the $f$-electrons are prevented from participating in bonding,~\cite{penicaud} 
indicating the need of a different energy functional.
For Pu and Am, situated at the localization-delocalization transition, neither of these two approaches
is capable of describing their electronic and magnetic properties, 
in particular conventional band theory finds both Am and $\delta$-Pu magnetic.
The onset of 
spin-polarization in band theory may be interpreted as a signature of
$f$-electron localization,\cite{HLS} albeit only close
to half-filling, and when combined with anti-ferromagnetic ordering and orbital
polarization\cite{soderlind1,soderlind,soderlind2} a quite accurate account of bonding energetics is
obtained. The predicted  magnetic ordering is however in disagreement 
with the experimentally observed Pauli enhanced paramagnetic ground state in $\delta$-Pu and Am. 
By allowing for a complete disorder of the local magnetic moments, 
Niklasson et al.\cite{Niklasson}
obtained a reasonably accurate description of bond lengths and bulk moduli throughout the actinide series,
however, the underlying assumption of fluctuating local moments has not been confirmed by experiments. 

The inadequacy of the local density functional for the later actinides can be traced to its inaccurate representation of
strong electron-electron interaction, which tends to localize the $f$-electrons on their atomic sites. 
To overcome this shortcoming,
all implementations of the LDA+$U$ approach\cite{bouchet,savrasov,shorikov,shick} 
introduce the effective
Coulomb $U$ parameter that separates the $f$-manifold into lower and upper Hubbard
bands, and removes $f$-degrees of freedom from the Fermi level. In the fully localized limit (FLL) LDA+U
finds $\delta$-Pu magnetic,\cite{savrasov,bouchet,shick} 
but 
a careful balancing of spin-orbit and exchange 
interactions in the
implementation of Ref. \onlinecite{shorikov} leads to a 
non-magnetic ground state of $\delta$-Pu. 
In the around mean-field (AMF) flavour of LDA+U non-magnetic ground states
were found for both $\delta$-Pu and Am.\cite{shick,shickagain}
The recent development of dynamical mean-field theory (DMFT),\cite{DMFT} combined with
the LDA+$U$ approach, has enriched the field in terms of phenomena accessible to calculations, 
including both ground state cohesion,\cite{savrasov1} phonons\cite{dai} and
photoemission.\cite{savrasov2,Landeretal,dmft}
However, most applications to date invoke the Hubbard Hamiltonian and inherit
the uncertainties associated with the LDA+$U$ method, most notably the effective $U$
and the double counting correction,
shortcomings that might be overcome if DMFT could be successfully 
merged with the GW technique.\cite{Silke}
The disordered local moment approach of Ref. \onlinecite{Niklasson} is
related to DMFT by a static approximation. 

The self-interaction corrected local spin density (SIC-LSD) theory\cite{SSC} provides a dual picture of
coexisting localized and
band-like $f$-electrons,
which was also conjectured\cite{olle-delta} to be the appropriate picture of the $\delta$-phase of Pu.
The method is fully ab-initio as both kinds of $f$-electrons are treated on an equal footing, with no adjustable 
parameters. Consequently, a unique total energy functional can be applied to the entire range of actinide elements
to determine the ground state localized/delocalized $f$-electron manifold of each element.
The method was previously applied to the actinide elements Pu to Cf assuming a
ferromagnetic ground state and, with the exception of Pu, the calculated volumes and bulk moduli were in 
reasonably good agreement with the experimental values.\cite{SSC}
The motivation for the present paper is to extend the SIC-LSD study to other magnetic orderings 
(para-, ferro- and antiferro-magnetism), 
and explore the consequences for the bonding properties and valence state of the actinides. 
In addition, the delocalization of $f$-electrons under pressure in the later actinides is also investigated. 
Through the present more comprehensive study, we seek to establish and document the SIC-LSD 
applicability and status regarding
a thorough physical understanding of such complex systems as actinides. 
%

The magnetic and paramagnetic states are 
realized by                                                
switching on or off the spin dependent part of the exchange-correlation potential.
We find that, except for U and Np, the spin-polarized solution has always a 
lower energy than the non-magnetic solution, which in case of Pu and Am is in disagreement with 
experimental observation. However, it emerges that assuming a paramagnetic representation of
the localized states provides a much improved description of the equilibrium volumes of both 
$\delta$-Pu and Am.
Antiferromagnetic order 
is found to be the lowest energy solution for both Cm and Bk, in agreement with experiment, providing
slightly improved specific volumes compared to the values calculated for the ferromagnetic 
case. Since SIC-LSD describes well the delocalized nature of the $f$-electrons in the early actinides and the 
localized nature of the $f$-electrons in the late actinides, this demonstrates that the transition from 
$f$-bonding to $f$-localization in the $5f$ series is reproduced, 
while an adequate description of the cross-over region 
depends strongly on the representation used for the localized states.

The paper is organized as follows. In Section II the important aspects of the SIC-LSD method are
outlined. In Section III we present our results for the U, Np, Pu, Am, Cm and Bk metals.
The observed trends of the present study are discussed in Section IV, and Section V gives our
conclusions.

\section{Theory}
       
\subsection{The SIC-LSD total energy method}

The LSD approximation to exchange and correlation introduces an unphysical interaction of the
electron with itself,~\cite{zp} which, though insignificant for extended band states, may lead to uncontrollable
errors in the description of atomic-like localized states, for example the $f$-electrons in the later actinides. 
The SIC-LSD method~\cite{brisbane,cerium,chemist} corrects for this 
self-interaction, by postulating a manifold of coexisting localized and delocalized $f$-electrons, and
by adding to the LSD total energy functional an explicit energy contribution for an electron to localize. 
The approach is fully ab-initio,  
as both localized and delocalized states are expanded in the same set of basis functions, 
and are thus treated on an equal footing.
One major advantage of the SIC-LSD energy functional 
is that it allows to realize and study different valence states
of ions constituting a solid. By assuming atomic configurations with different total numbers of
localized states, self-consistent minimization of the total energy leads to different local
minima of the same total energy functional, $E^{SIC-LSD}$, 
and hence their total energies may be compared. The
configuration with the lowest energy defines the ground state configuration. 
If no localized states are assumed, $E^{SIC-LSD}$ coincides with the
conventional LSD functional, {\it i.e.}, the Kohn-Sham minimum of the $E^{LSD}$
functional is also a local minimum of $E^{SIC-LSD}$. 
Another advantage of the SIC-LSD scheme is the possibility to localize $f$-states of 
different symmetry.  In the present work we exploit this possibility to investigate 
several magnetic structures including non-magnetic, ferromagnetic and antiferromagnetic order.

Given the total energy functional $E^{SIC-LSD}$, the computational procedure is as for the LSD case,
{\it i.e.} minimization is accomplished by iteration until self-consistency.
In the present work, the electron wavefunctions are expanded in 
linear-muffin-tin-orbital (LMTO) basis functions,\cite{OKA}
 and the energy minimization problem becomes a
non-linear optimization problem in the expansion coefficients, which is only slightly more
complicated for the SIC-LSD functional.
Further details of the present implementation
can be found in Ref. \onlinecite{brisbane}.
For a variety of applications to $d$ and $f$ electron solids, see Refs.
\onlinecite{brisbane,nature,pss,pnictides,rules,puo2} and references therein.

The elementary excitations of solids, i.e. the one-particle energies, are often approximated by
the band eigenenergies of the Kohn-Sham scheme, although a formal justification does not
exist. 
Although for weakly correlated materials such an approximation 
may by reasonable, for localized electron states it is ill founded.
It is possible to define band states constructed as Bloch combinations of
the localized states (sometimes referred to as 'canonical orbitals'\cite{heaton}), however the
associated band energies bear little resemblance to physical removal energies, as e.g.
observed in photoemission spectroscopy. 
This is in contrast to the LDA+$U$ approach where better agreement with photoemission experiment
can be accomplished by an appropriate 
choice of the $U$ parameter and double-counting correction. On the other hand, the SIC-LSD method,
being fully ab-initio, does not require parameters, and can therefore be applied to study trends
in energetics of the entire series of actinides in a comprehensive manner, and with potentially
predictive capability.
The SIC-LSD method is not a means to explain spectroscopies but should be treated as a useful
total energy approach, allowing to study a variety of phenomena and among them valences and valence
transitions.

\subsection{Calculational Details}

Calculations have been performed for 
the face-centered cubic (fcc), hexagonal close-packed (hcp), and
double hexagonal close-packed (dhcp) crystal structures. 
The Wannier states describing the localized $f$ electrons are expanded on clusters of 32, 64 or
48 sites around the localization site,\cite{cerium} for these three cases.
Similarly, the itinerant states have been  sampled using 525, 216 or 75
k-points in the irreducible wedge of the Brillouin zone, respectively.
The $6s$ and $6p$ semicore states have been treated self-consistently in a separate energy panel.
In the valence panel the LMTO basis set includes  functions of $7s$, $7p$, $6d$ and $5f$
character, where the $7p$ channel has been downfolded.\cite{OKA}

The atomic spheres approximation (ASA) is used, whereby the crystal volume is divided into 
slightly overlapping atom-centered spheres of a total volume equal to the actual volume.
The ASA gives rise to uncontrollable errors in the evaluation of the total energy, which inhibits
the comparison of energetics of different crystal structures, on account of different overlaps
of the atomic spheres. In the present study the
focus is on the energetics of different localization scenarios within the same crystal
structure, and the ASA error is of minor influence.  
The spin-orbit interaction couples the band Hamiltonian for the spin-up and spin-down channels, 
{\it i.e.}  a doubled secular problem must     be solved.
Other relativistic effects are automatically included by solving the scalar-relativistic
radial equation inside spheres.
The spin-orbit parameter, 
\[
\xi(r)=-\frac{2}{c^2}\frac{dV}{dr},
\]
 in atomic Rydberg units, 
is calculated from the self-consistent potential.

 

\begin{table*}
\label{magn}
\caption{Equilibrium volumes, $f$-occupation and magnetic moments of U, Np,
$\delta$-Pu, Am, Cm, and Bk in the SIC-LSD ground state.
Results for both the paramagnetic and the magnetic ground state are quoted.
The numbers of localized and delocalized $f$-electrons are given, and for the
moments both the spin and orbital moments (in Bohr magnetons, $\mu_B$) are listed. All paramagnetic (PM)
calculations were done for the fcc structure, while ferromagnetic (FM) calculations were
performed for the fcc structure in the case of U and Np,
and antiferromagnetic (AFM) calculations for the hcp structure for Pu, Am, Cm, and Bk.
Last column gives the energy difference (in eV) between the paramagnetic and magnetic ground
states.}
\begin{ruledtabular}
\begin{tabular}{|l|cc|cc|c|c|c|}
     &   $V_0$ (a.u.) &$V_{exp}$(a.u.)&$N_f$(loc) &  $N_f$(band)  &  $M_L$   &    $2M_S$   &
              $\Delta E_{PM-AFM}$ (eV) \\
 \hline
    U       &  140.4  &    138.9   &     0        &      3.07     &  0.0     &    0.0  &  -  \\
\hline
    Np  &  122.4  &    126.9   &     0        &      4.30     &  0.0     &    0.0  &  -  \\
\hline 
    Pu (PM) &  163.8  &         &     4        &      1.34     &  0.0     &    0.0  &    \\     
   Pu (AFM)  &  207.1  &    \raisebox{1.5ex}[-1.5ex]{168}        &     5        &      0.35     & -4.11    &   5.25  &  \raisebox{1.5ex}[-1.5ex]{$1.14$}    \\
\hline 
    Am (PM) &  200.6  &         &     6        &      0.40     &  0.0     &    0.0  &    \\    
    Am (AFM) &  210.5  &    \raisebox{1.5ex}[-1.5ex]{198}       &     6        &      0.54     & -1.59    &   6.45  &  \raisebox{1.5ex}[-1.5ex]{$1.56$}    \\
\hline 
    Cm (PM) &  171.0  &         &     6        &      1.51     &  0       &    0    &    \\   
    Cm (AFM) &  203.8  &     \raisebox{1.5ex}[-1.5ex]{202}     &     7        &      0.26     & -0.15    &    6.79 &  \raisebox{1.5ex}[-1.5ex]{$2.29$}     \\
\hline 
    Bk (PM) &  205.1  &        &     8        &      0.67     &  0       &   0     &    \\   
    Bk (AFM) &  197.4  &    \raisebox{1.5ex}[-1.5ex]{189}     &     8        &      0.43     &  3.21    &    5.36 &  \raisebox{1.5ex}[-1.5ex]{$1.86$}     \\
\end{tabular}
\end{ruledtabular}
\end{table*}

Given the considerable disagreement between theory and experiment as to the magnetic ordering
specifically in Pu and Am, we use the total energy considerations to investigate the
SIC-LSD ground state with respect to both valency and spin configuration.
The magnetic order is dictated by the order imposed on the localized $f$-states, and
the paramagnetic and ferromagnetic/anti-ferromagnetic scenarios are implemented 
by considering the initial localized $f$-manifold to be
 either non-spin-polarized or spin-polarized in appropriate spatial arrangements.\cite{MXRS}
During iteration to self-consistency
the Wannier states may change, however roughly retaining
their overall characteristics (exceptions occur
close to delocalization).

Other approaches implement a non-magnetic ground state either by assumption (DMFT),\cite{savrasov1}
through a careful calibration of the $U$ and $J$ parameters in LDA+U (FLL)\cite{shorikov} or the
double counting correction in LDA+U (AMF),\cite{shick}
or by invoking disorder in the directions of local moments.\cite{Niklasson}

\section{Results}

As described in the previous section, we consider in this work only
the high symmetry crystal structures relevant
for the actinide elements in their low-pressure/large specific volume phases 
with localized $f$-electrons, specifically 
fcc and hcp crystal structures. A few tests
for the dhcp structure, which is the actual ground state structure     at
ambient conditions for Am, Cm and Bk, revealed virtually no difference with respect to the
simpler hcp structure in terms of cohesive energy per atom, specific volume and magnetic
moments. As mentioned earlier, the present ASA implementation of the SIC-LSD energy functional is not 
accurate enough to resolve the energy differences between these structures.

\subsection{Total energy considerations}

In Table I, the results of our calculations are summarized. In the last column, the calculated total energy
differences between the paramagnetic and magnetic configurations are shown.
We find that the non-magnetic configuration is energetically most favourable 
for the light actinides, U and Np, while all the remaining actinides prefer the spin-polarized ground state.
Thus, from SIC-LSD total energy considerations we find the calculated ground states in agreement with experiment
for U, Np, Cm, and Bk, but fail to reproduce the experimentally observed non-magnetic ground states of both 
Pu and Am.

The fact that total energy considerations in LSD and also
SIC-LSD favour the magnetic     ground state in the actinide elements Pu and Am 
is a manifestation of the Hund's first rule rather than a consequence of interatomic exchange being significant.
Neither LSD nor SIC-LSD can describe the
appropriate multi-determinant eigenstates of $L$ and $S$. This leads to an error
in the energy associated with local paramagnetic moments, since
the exchange potential generated by the localized states and exerted on the non-$f$
electrons is determined by $M_S$ ($\sim S$). 
This is most clearly demonstrated  for $f^6$, where the correct ground state according to
the Hund's rules in Russell-Saunders coupling is $^7F_0$, i.e. $S=L=3$, $J=0$.
Hence, also
$M_S=0$, and accordingly  no spin-dependent exchange-field is exerted on the conduction
electrons by an actinide ion in the $f^6(^7F_0)$ configuration.
But LSD sets up a large spin-dependent potential, since it can only represent the
$^7F_0$ ground state by $M_S=3$.
This description can be valid if the inter-atomic exchange
interaction is strong enough for an ordered magnetic phase to be formed.
However, it is an incorrect description of the cohesion of the conduction
states, if the exchange interaction is not strong enough compared to the
spin-orbit interaction which selects the $J = 0$ ground state multiplet.
The latter scenario occurs for Am. This also explains why the disordered
local moment approach describes the actinides so well,\cite{Niklasson} since the disorder
(in the direction of spin quantization axis) leads to  averaging
of the spin-dependent part of the potential to zero.


\subsection{Uranium and Neptunium}

Studies of the electronic structure of U and Np, have demonstrated
that the observed volumes and crystal structure can be explained by the conventional band structure
calculations (see for example Ref. \onlinecite{singh} and references therein). 
Nevertheless, despite this overall rather good agreement with experiment there 
are still discussions as to what 
band structure approach is best suited to describe volume trends.~\cite{singh,soeder,Nordstroem} 
Furthermore, 
a comparison between XPS data
and a fully relativistic calculation has shown that correlations already have a measurable influence on 
the electronic structure of Np.~\cite{NEC}
\begin{figure}
\begin{center}
\includegraphics[width=90mm,clip]{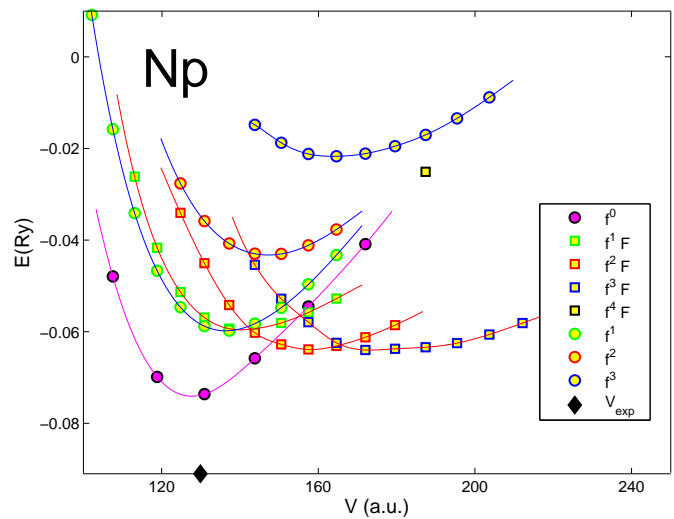}\\
\caption{
\label{Npenergy}
(Color online)
Total energy of fcc Np as function of volume for several
scenarios distinguished by the different 
number of localized $f$-electrons on Np ions.
Both non-magnetic (spheres) and ferromagnetic (squares) configurations are considered.
}
\end{center}
\end{figure}

In our calculations,
we compare the band-$f$ scenario, where all $f$ electrons are treated as itinerant, to
a number of localized/delocalized $f$-electron scenarios, and find it 
to be energetically most favourable for both U and Np. 
In figure \ref{Npenergy}, the total energy versus volume is plotted for a number of Np $f$-electron configurations,
both for the non-magnetic and the magnetic states. The global total energy minimum is obtained
for the $f^0$ configuration.
Spin polarization is not energetically favourable, resulting in a  non-magnetic ground state 
in agreement with experimental data for 
both U and Np. This indicates that correlations are
relatively weak in these elements, and confirms the adequacy of the
LSD picture for treating the electronic structure of the early actinides.
Comparing the calculated equilibrium volumes to the experimental values, we find very good agreement for both
U and Np, but with some slight overbinding in Np, as is also observed in other band structure calculations. 
A comparison of the difference in total energies between the LSD ground state configuration and the nearest
localized configuration gives respectively 0.38 eV for U and 0.20 eV for Np, implying that the localized
configuration is less unfavourable for Np, which indicates  the increasing influence of correlations.

\subsection{Plutonium}

From the theory point of view, Pu is without doubt the most interesting element of 
the actinide series, as witnessed
by the considerable number of papers already published on the subject.   
Its complex phase diagram, behaviour under pressure, and the absence of magnetism have still
not been fully understood to date.
Within DFT a number of theoretical studies have concentrated on the magnetic 
ground state, where spin-polarization mimics the effect of $f$-electron localization,
which leads to calculated lattice parameters that are in rather good agreement with the experimental values, 
for both
the $\alpha$- and $\delta$-phase.~\cite{soderlind2,bouchet,kutepova} 
Other calculations have tried to take the experimentally observed non-magnetic ground state into account,
either by assuming it as a starting point of their calculation~\cite{Niklasson,savrasov1} 
or by implementing it into their model by means of an adequate choice of parameters.~\cite{savrasov,shorikov,shick}
Using the latter approach, LDA+$U$ calculations\cite{shorikov,shick} find a non-magnetic
ground state for metallic Pu in both phases, driven by a strong spin-orbit coupling in the 5$f$-shell.

\begin{figure}
\begin{center}
\includegraphics[width=90mm,clip]{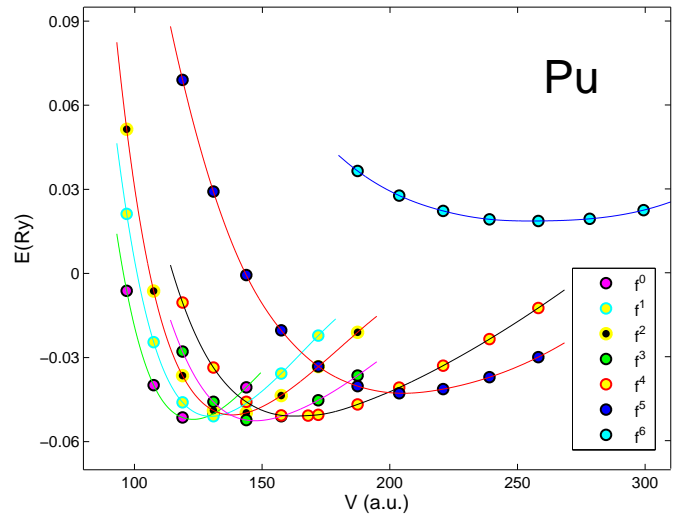}\\
\caption{
\label{Pujj}
(Color online)
Total energy of paramagnetic fcc Pu as function of volume for several
scenarios distinguished by the different 
number of localized $f$-electrons on Pu ions.
}
\end{center}
\end{figure}

In our calculations for Pu we merge elements of both of these approaches,
as in what follows we assume a non-magnetic ground state, 
and compare to our previously published ferromagnetic calculations.\cite{SSC}
In figure \ref{Pujj}, the
total energy of Pu in the fcc phase is displayed for several localized
paramagnetic $f^n$ configurations. 
Remarkably, the energy minima for the scenarios with 0, 1, 2, 3 or 4 localized
states are nearly degenerate (within 0.03 eV/atom), while 
localizing 5 $f$-electrons is marginally less favorable (0.11 eV/atom higher energy than the $f^4$
scenario) and 
localizing 6 $f$-electrons is 0.90 eV/atom higher in energy. The implication is that 
fcc Pu is situated virtually at the localization-delocalization transition at ambient pressure.
The Pu $f$-manifold is dynamical, as the $f$-electrons may freely
fluctuate between a localized $f^4$ shell and a fully
itinerant state. It signals an electronic wavefunction of
much more complex nature 
than can be realized with the SIC-LSD approach, which in the usual spirit of DFT 
seeks to describe the many-electron system with a single Slater determinant. 
The equilibrium volume of the $f^4$
state is calculated as 163.8 a.u., which is close to the experimental value for the 
$\delta$-Pu volume of 168 a.u., while the calculated
equilibrium volume of the $f^0$ state, 123.0 a.u.,
is  close to, but smaller than,
the experimental equilibrium volume of $\alpha$-Pu, 135 a.u., quite a
typical overbinding for the LSD
approximation.
Hence, it is suggestive that the $f^4$ phase be stabilized by high temperature or alloying~\cite{baclet}
to form the $\delta$-phase of Pu, while the low symmetry crystal structure of the $\alpha$-phase
most probably stabilizes the $f^0$ scenario.

\begin{figure}
\begin{center}
\includegraphics[width=70mm,clip]{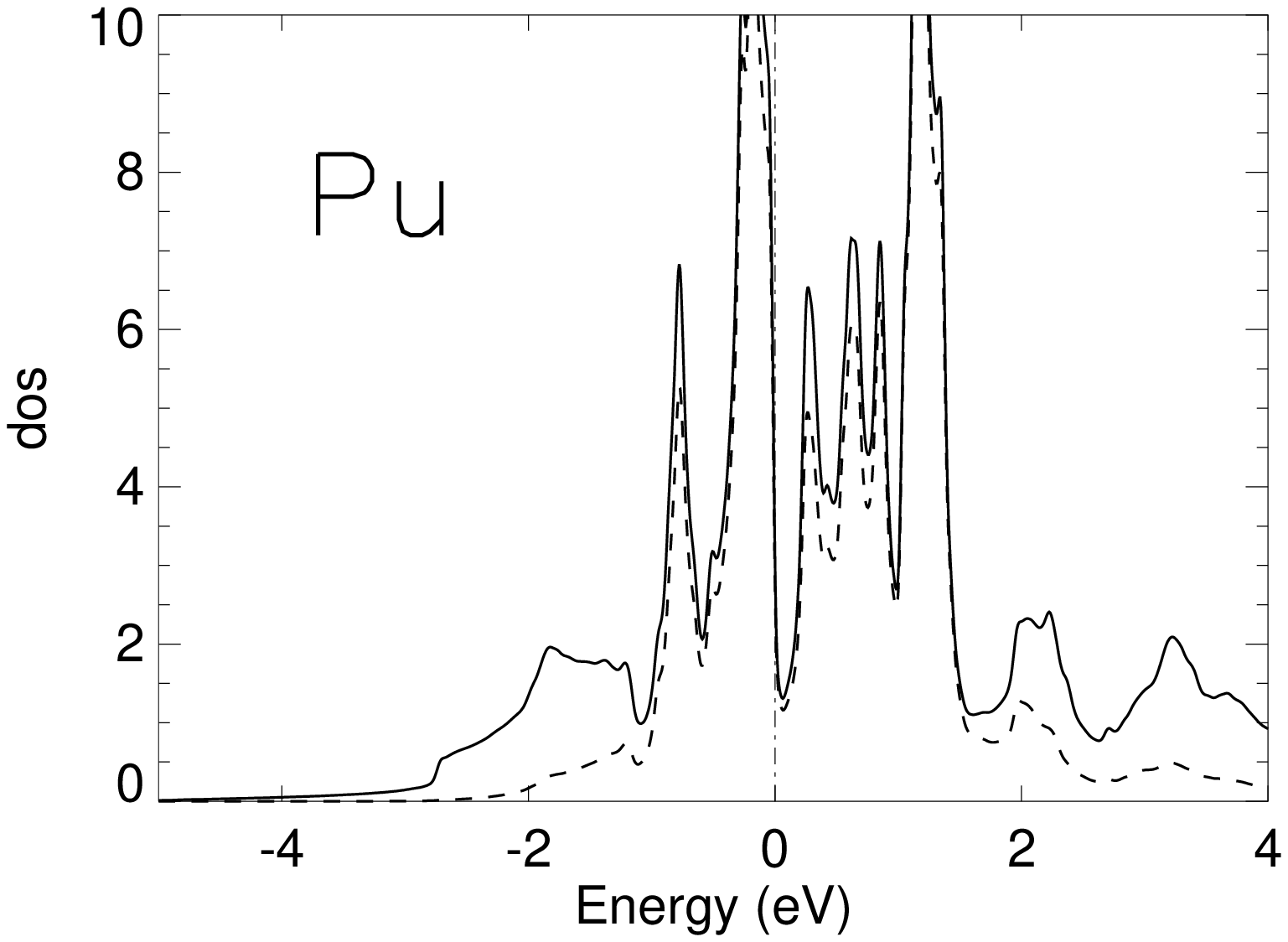}
\includegraphics[width=70mm,clip]{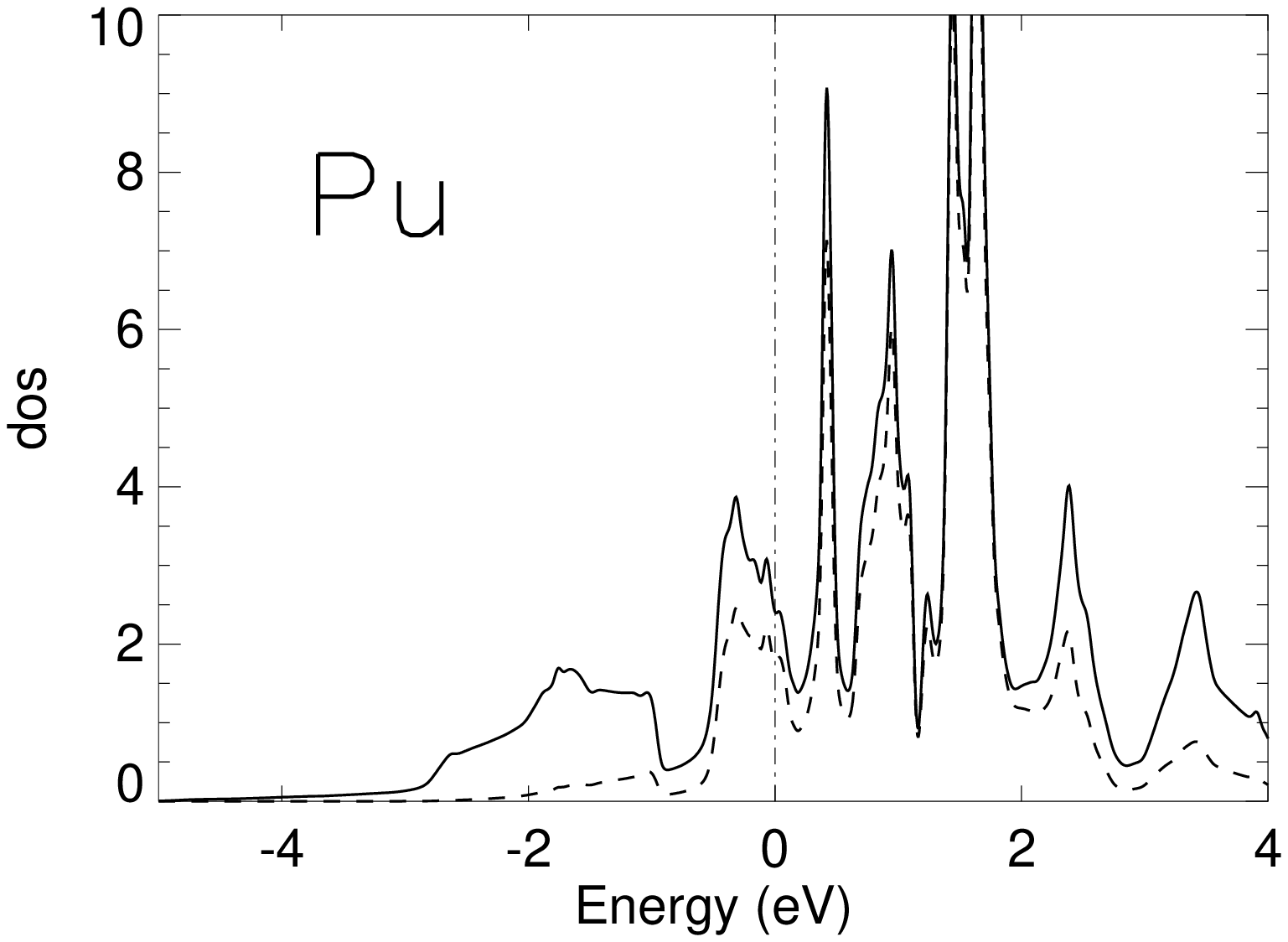}
\includegraphics[width=70mm,clip]{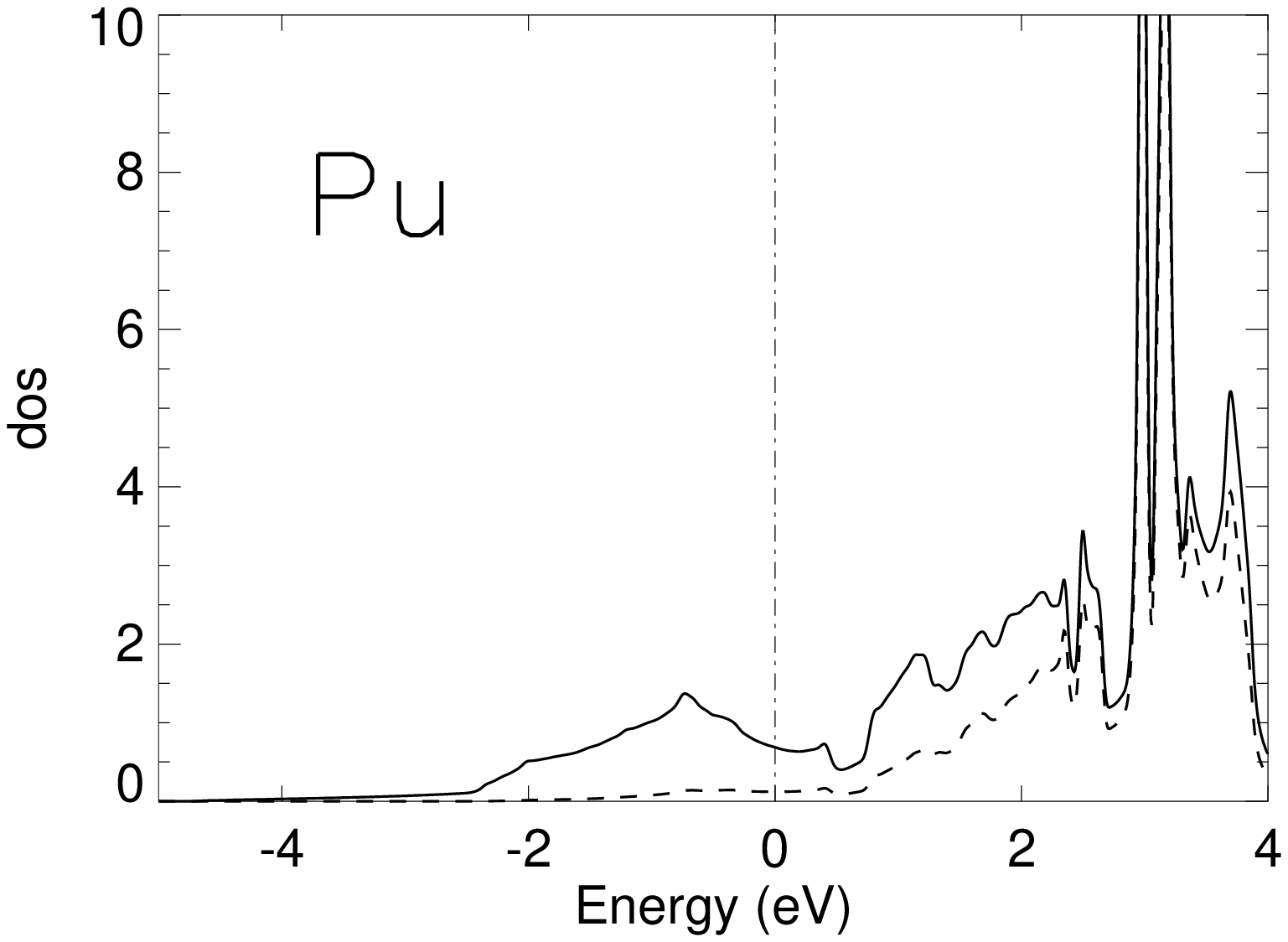}
\caption{
\label{PuDOS}
Density of states of paramagnetic fcc plutonium in a) the LSD scenario b) the localized $f^4$ scenario
and c) the localized $f^6$ scenario. 
Full line is the total density of states, while dashed line gives the $f$ partial
density of states. Units are states per atom and per eV, and the energy is given relative to
the Fermi level.
}
\end{center}
\end{figure}

In figure \ref{PuDOS} the density of states (DOS) of a number of these localization scenarios 
is displayed.
In the LSD scenario of figure \ref{PuDOS}a, the $f$-manifold is
split into a $j=5/2$ and a $j=7/2$ band by the spin-orbit interaction. The exchange interaction
is absent  here, and the additional splitting within each of the two groups of bands is
due to crystal field effects. Here the Fermi level is situated at the top of the filled $j=5/2$ band, with
a resulting large DOS at the Fermi level. 
In Fig. \ref{PuDOS}b, which represent the DOS of the $f^4$ paramagnetic configuration,
four of the $f$-states treated as band states in Fig. \ref{PuDOS}a, 
have become localized, leaving the remaining $f$-states available for
band-formation.
The width of the valence bands is 5 eV, however with most of the weight in the region from
the Fermi level to 2.5 eV below the Fermi energy. A considerable hybridization of band
$f$-states is  seen in the range from -1 eV below to the Fermi level, including a high density
of states at the Fermi level, in 
agreement with the experimental photoemission
spectrum\cite{pu-pes} and the high specific heat\cite{pu-cv} of $\delta$-Pu. 
We note however that a photoemission spectrum is not directly
comparable to a theoretical density of states, which strictly speaking only provides 
information on how the ground state is constructed from one-particle states,
while the final excited states play a crucial role in the photoemission spectrum.

In the earlier SIC-LSD calculation\cite{SSC}, assuming a ferromagnetic ordering, we found
a Pu ground state comprised of localized $f^5$ ions, with much too large a volume
(30 \% overestimate), quite like the paramagnetic $f^5$ scenario in figure \ref{Pujj},
which has a minimum volume
around 205 a.u.. 
By invoking ferromagnetic order in the LDA+U (FLL) calculations, Refs. \onlinecite{savrasov,bouchet,shick}
also concluded on an $f^5$ localized ground state. The recent LDA+$U$ calculations 
of Ref. \onlinecite{shorikov} based on
a reduced exchange interaction determined a non-magnetic
ground state characterixed by $f^6$ ions, with
a vanishing DOS at the Fermi level. A similar picture emerges from our SIC-LSD
DOS for the $f^6$ configuration, shown in Fig. \ref{PuDOS}c. 
Energetically we find this configuration unfavourable by almost 1 eV, and the absence of any noticeable DOS at
the Fermi level is in disagreement with experiment. 
The AMF flavour of LDA+U finds $\delta$-Pu in a mixed valent state\cite{shick} with a total $f$-occupancy of 5.44.
In a recent DMFT calculation,~\cite{shim} Pu is similarly described as being in a mixed valence state
(strong peak around
the $f^5$ configuration), with an average $f$-occupation of 5.2, while another DMFT calculation
using a different impurity solver suggest a value around 5.8.\cite{Landeretal}
The integrated number of band $f$-states in our $f^4$ configuration is 1.34, 
therefore the total number of $f$-electrons in $\delta$-Pu comes out to be 5.34 (see Table I). 
This result is markedly different from the situation in the rare earths where 
the number of delocalized $f$-electrons never exceeds about $0.7$,\cite{nature}
{\it i. e.} significantly more 
$f$-electrons are involved in the cohesion in $\delta$-Pu at ambient conditions.
The total number of $f$-electrons is  nearly a constant in the various localization scenarios 
$f^0$ - $f^5$ ($N_f= 5.48, 5.42, 5.36, 5.36, 5.34, 5.39,$ respectively), 
{\it i.e.} what distinguishes these cases is the ratio of
localized to delocalized $f$-electrons. The $f^6$ scenario becomes energetically unfavorable
(figure \ref{Pujj}) precisely because it forces a high $f$-occupancy on Pu ($N_f=5.93$, 
slightly below 6 due to the
tails of localized states reaching into the neighbouring sites), 
which involves a relatively high intraatomic Coulomb interaction energy.

The localized $f$-states are not shown in figure \ref{PuDOS}, since 
their SIC band energies are poor estimates of physical removal energies.
A rough estimate of removal energies of the localized states may be obtained by
the transition state argument,~\cite{dms} which when applied to the $f^4$ 
state of Pu places the localized states at -3.1 eV. In experiments,\cite{pu-pes}
there is no evidence of particularly strong emission at this energy, so this aspect of
the electronic structure of $\delta$-Pu remains a puzzle.

In conclusion,
it emerges that the paramagnetic SIC-LSD implementation describes the fcc phase
of Pu satisfactorily, while the combination of the self-interaction correction and the
exchange energy in the magnetic limit  overestimates the tendency to localization.
Experimentally, the vanishing of magnetic moments in Pu was recently re-confirmed.\cite{lashley}
Furthermore, electron energy loss spectroscopy branching ratios\cite{tobin,tobin1}
 clearly favor the $jj$-coupling picture over the $LS$-coupling (for $\alpha$-Pu).

\begin{figure}
\begin{center}
\includegraphics[width=90mm,clip]{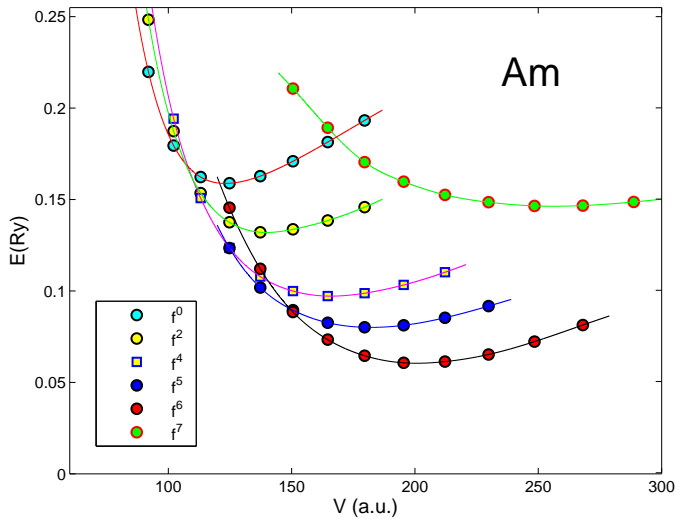}\\
\caption{ 
\label{Amjj}
(Color online)
Total energy of paramagnetic 
fcc Am as function of volume. 
 Several paramagnetic configurations of the localized $5f$ subshell are considered, 
as discussed in text.
}
\end{center}
\end{figure}

\begin{figure}
\begin{center}
\includegraphics[width=90mm,clip]{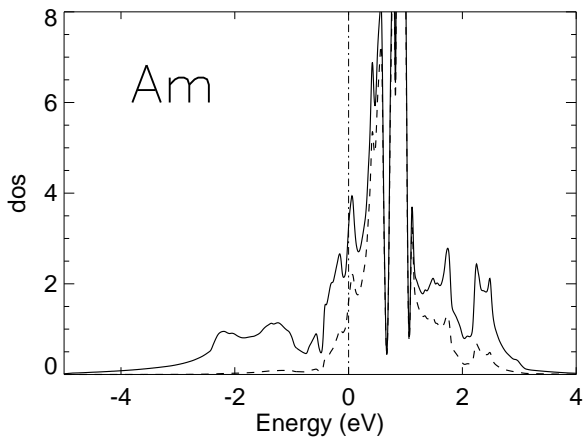}\\
\caption{ 
\label{Amdos}
Density of states of hcp americium in the paramagnetic
state with 6 localized $f$-electrons
per Am atom at $V=V_0=200.6$ a.u.. Full 
line is the total density of states, while dashed line gives the $f$ partial
density of states. Units are as in figure 3.
Only the itinerant electron states are shown. The localized states by the transition state
argument may be estimated to be situated around -4.0 eV.
}
\end{center}
\end{figure}

\subsection{Americium}

The SIC-LSD total energy curves for americium are displayed in figure \ref{Amjj}. As for Pu we
have investigated several localization scenarios from
$f^0$ to $f^7$, assuming a paramagnetic phase. 
The lowest energy is found for the $f^6$ configuration with an equilibrium volume of $V_0=200.6$
a.u. per atom, which matches perfectly with the experimental volume of $V_{exp}=198$ a.u..
At this volume the $f^5$ configuration is 0.27 eV higher, and the $f^7$ configuration is 1.4 eV 
higher, i.e. the $f^6$ ground state appears quite robust at ambient conditions, contrary to the
situation in Pu. 
This result is in agreement with recent photoemission studies~\cite{gouder} 
that measure a solid $f^6$ ground state
configuration.
A non-magnetic $f^6$ ground state configuration has also been suggested 
on the basis of LDA+$U$ (AMF) calculations.~\cite{shickagain}

The density of band states of Am in the paramagneitc $f^6$ ground state is displayed in Fig. \ref{Amdos}.
The calculation shown has been performed
for the hcp closed packed 
structure, but the 
density of states  of the fcc
structure is very similar. The Am $f$-bands are almost exclusively of $j=7/2$ character and
situated around 1 eV above the Fermi level, with a small tail extending below the Fermi level.
The total number of $f$-electrons is $N_f=6.40$, i.e. approximately 0.4 $f$-electrons reside in
this tail.
The localized states are not shown in the figure but the transition state estimate places them
at an energy $\sim -4.0$ eV. Compared to $\delta$-Pu, Am has 
approximately one less  band $f$ electron, slightly
deeper localized states and more narrow unoccupied $f$-bands. Early photoemission experiments\cite{naegele} 
on Am reveal the localized states as a plateau 
between 2 and 4 eV below the Fermi level, with traces of $f^7$ character\cite{borje2,dmft}
between -2 and -1 eV, however with virtually no $f$-related emission at the Fermi level.
Hence the electronic structure of Am is very different from that of $\delta$-Pu. The
photoemission spectra have been interpreted in terms of a mixed valent ground state
of fluctuating $f^6$ and $f^7$ configurations,\cite{dmft} dominated by the former. The present
SIC-LSD ground state, comprised of firm $f^6$ ions coexisting with a tail of 0.4 band $f$ electrons,
can be considered a reasonable representation of the same mixed valent ground state.

Upon compression the Am $f$-shell becomes more and more active in the bonding, as illustrated
in figure \ref{Amjj}, where the energy curves of the configurations with fewer
localized/more delocalized
$f$-electrons come closer to the $f^6$ curve. At a compression around $V/V_0=0.75$
the $f^5$ curve  becomes lower in energy, around $V/V_0=0.65$ the $f^4$ curve dips
below and becomes most favorable and so on, until around $V/V_0=0.52$ eventually the
$f^0$ curve becomes the lowest, meaning that the lowest energy is obtained with complete band-description
of the $f$-electrons. Hence the $f$-delocalization happens gradually in the volume range between
$V/V_0\equiv 0.75$ and $V/V_0\equiv 0.52.$ In terms
of pressure, this corresponds to the range
$p=16$ GPa  to  $p=35$ GPa, as given by the common tangents of the respective energy curves. Of
course, since the crystal structure in figure \ref{Amjj} is fixed to be
fcc this is not directly
comparable to the experimental behavior of Am under pressure,\cite{heathman} where the 
$f$-electron delocalization is accompanied by a sequence of structural changes. However, if we
assume that the AmI (dhcp) and AmII (fcc) phases are indicative of localized $f$-electrons, the
experimental onset of delocalization is at the AmII$-$AmIII phase transition, which occurs at
$V/V_0=0.77$ and $p=10.0 $ GPa.  The $f$-electron
delocalization is presumably  nearly
complete in the AmIV (primitive
orthorhombic), which is obtained at $V/V_0=0.63$ and $p=17.5$ GPa. This volume is larger than
the $V/V_0=0.52$ deduced from theory, most probably due to the different crystal structure, as
it is well known that $f$-electron delocalization favors lower symmetry
lattices\cite{ErikssonNature} and vice versa. The interpretation of the AmIII (face centered
orthorhombic) as a structure of intermediate $f$-electron delocalization is also corroborated by
total energy calculations,\cite{soderlind} for which additional $f$-correlations (in the form
of anti-ferromagnetic spin-polarization and orbital polarization) need to be
incorporated to achieve accurate description.
Hence, there is
qualitative agreement between the SIC-LSD theory of $f$ electron delocalization under pressure
and experimental observations for americium.

In our earlier SIC-LSD study of americium\cite{SSC} 
only the ferromagnetic spin arrangement was considered. The calculated ground state
configuration was also $f^6$, however with an 8 \% too large equilibrium volume, and the $f^7$
configuration only 0.12 eV higher in energy. In addition, the spin and orbital moments were far from
cancelling ($M_S=3.41 \mu_B$, $M_L=-1.58 \mu_B$, wher $\mu_B$ is a Bohr magneton). The situation is 
slightly improved by allowing for an antiferromagnetic spin arrangement. In the hcp structure, the 
energy per atom is lowered by 0.03 eV in the
antiferromagnetic structure compared to the ferromagnetic arrangement, 
and the equilibrium volume is lowered by 5 \%, but the moments remain
almost unchanged ($M_S=3.23 \mu_B$, $M_L=-1.59 \mu_B$). Hence, 
antiferromagnetic ordering constitutes the lowest energy solution of the SIC-LSD minimization.
Overall, however, the paramagnetic scenario provides the best description of the
elemental americium, while imposing an antiferromagnetic order can only be considered as 
a simple means of partially patching up the bonding properties.
A SIC-LSD study on Am pnictides and chalcogenides predicted large DOS at the Fermi level, 
in agreement with susceptibility measurements~\cite{kanellak}, but not observed in photoemission
studies.~\cite{gouder} The results were obtained based on the assumption of a magnetic ordering, 
but invoking instead a paramagnetic state 
might provide some new insight into their electronic structure.

\subsection{Curium}

\begin{figure}
\begin{center}
\includegraphics[width=90mm,clip]{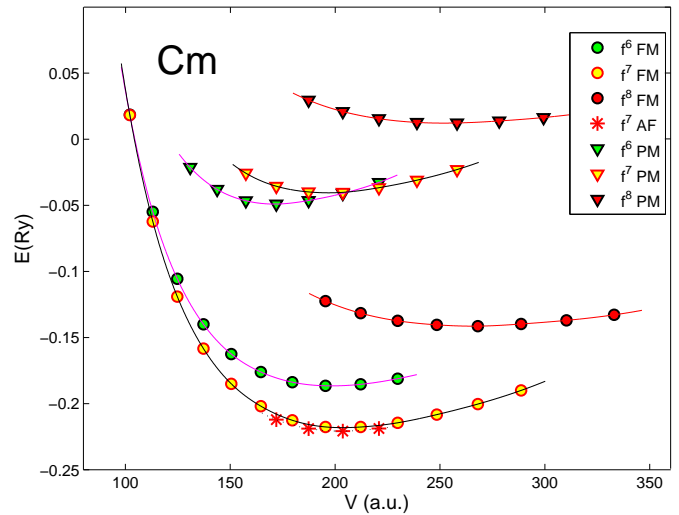}\\
\caption{
\label{CmLS}
(Color online)
Total energy of fcc Cm as function of volume for several scenarios differing with respect to the
number of localized $f$-electrons on Cm, notably 6, 7, or 8 localized $f$-electrons.
The balls correspond to the ferromagnetic ordering scenario, 
while triangles correspond to the non-magnetic scenario, as discussed in text. In addition, the
asterisks correspond to hcp antiferromagnetic arrangement of $f^7$ curium atoms.
}
\end{center}
\end{figure}
The SIC-LSD total energy of curium is shown for both the magnetic and paramagnetic 
scenarios in figure \ref{CmLS}.
For simplicity, only the fcc structure is considered, except for a few calculations performed with
antiferromagnetic spin arrangement in the hcp structure (asterisks in figure \ref{CmLS}). The
half-filled 
$f^7$ configuration of localized states is seen to give the minimum, at
a volume of $V=203.8$ a.u.,
which is in good agreement with the experimental equilibrium volume.
From Table I, and figure \ref{CmLS}, we find the paramagnetic scenario
unfavourable in energy with respect to the anti-ferromagnetic scenario by 2.3 eV.
Here the magnetic state with localized $f^7$ ions having $S=7/2$, $L=0$, 
and $J=7/2$, provides an adequate description of Cm.
The paramagnetic state 
in this
case gives a poorer description, 
with $f^6$ localized $f$ shells and  a 17 \% too small
volume. The antiferromagnetic state is lower than the ferromagnetic by 0.04 eV per atom in the hcp
structure, but we are not able to resolve the structural energy difference between the very
similar fcc, hcp and dhcp phases. The antiferromagnetic phase is also observed
experimentally.\cite{huray} A recent DMFT~\cite{shim} study has similarly established the $f^7$ atomic configuration
for Cm.

The relative stability of the half-filled $f^7$ shell is evidenced by the high compression
($V/V_0\sim 0.60$) needed for  the next delocalization step, $f^6$, to be of comparable energy.
This is also in accord with the higher pressure needed experimentally to convert Cm into a low
symmetry structure indicative of $f$-contribution to bonding (CmIII monoclinic structure),
appearing at $p=37$ GPa and $V/V_0=0.65$.\cite{landerCm}

\subsection{Berkelium}

\begin{figure}
\begin{center}
\includegraphics[width=90mm,clip]{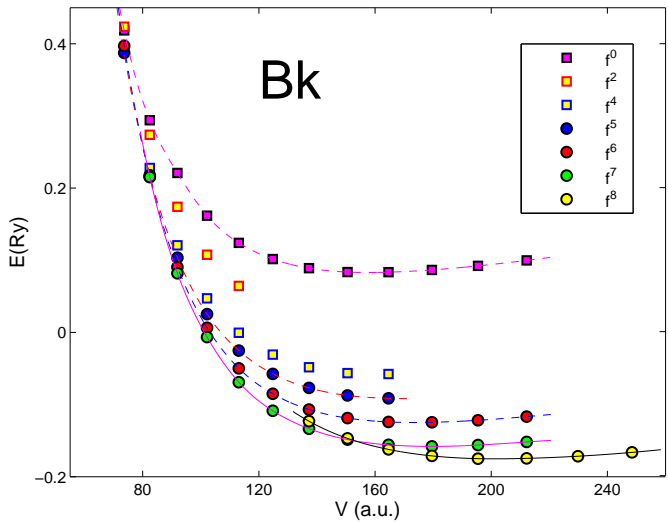}\\
\caption{ 
\label{BkLS}
(Color online)
Total energy of ferromagnetic
fcc Bk as function of volume for several scenarios differing with respect to the
number of localized $f$-electrons on Bk.
}
\end{center}
\end{figure}

The SIC-LSD total energies as a function of volume, calculated for berkelium in the fcc structure and the 
ferromagnetic spin-alignment, are shown in figure \ref{BkLS}. 
The localization $f^n$ scenarios, with $n=0,1,..,8$, have been considered, for both magnetic and paramagnetic 
states, but only the magnetic configurations are shown in the figure. 
The ground state is found for an $f^8$ localized shell, at a volume $V=201.4$ a.u., which is
6 \% larger than the experimental equilibrium volume. 
The energy difference between the ferromagnetic
and antiferromagnetic arrangements in the hcp structure is quite small, $\sim 0.02$ eV per atom,
 in favor of the latter.
The equilibrium volume at the same time decreases to 197.4 a.u.,
which is a small improvement over the ferromagnetic volume. 
The number of band $f$-electrons is 
0.43.
The magnetic moments of the Bk atom are large, $M_L=3.21 \mu_B$, $2M_S=5.36 \mu_B$, 
leading to a total moment of $M=8.57 \mu_B$. The experimental moment of
Bk is even higher, $\mu_{\mbox{eff}}=9.8 \mu_B$ 
(Ref. \onlinecite{huray}), leading to a projected value 
$M\sim 9.05$, i.e. very close to that of a pure $f^8$ Hund's rule coupled ion (here we assumed a
$g$-factor of 1.5 as appropriate for $f^8$).

The configuration with an $f^7$ localized shell is 
0.23 eV higher in energy. This state has 1.45 band-like
$f$-electrons. Upon compression this scenario becomes progressively more important,
and at $V/V_0=0.75$ it becomes lower in energy. The high stability of the $f^7$ shell is 
manifested through
this remaining the lowest energy scenario up to compressions of $V/V_0=0.40$, after which the $f$ manifold
quickly delocalizes. 
The transition pressure, corresponding to the
common tangent between the $f^8$ and $f^7$ total energy curves, is $p=    15$ GPa. Experimentally,
Bk is observed in the dhcp phase under pressures up to $p=7-8$ GPa, after which the fcc phase occurs and
is stable up to $p=25$ GPa and $V/V_0=0.70$.\cite{benedict} 
At pressures above 25 GPa
a low symmetry phase, possibly of $\alpha$-U type, is observed.
The conventional interpretation\cite{borje} is that the first two highly symmetric
crystal structures are associated with localized $f$-electrons, and hence that $f$-delocalization
sets in at 25 GPa, which is somewhat higher than the theoretical value found
here, but acceptable within 
our limits of accuracy.  It would be interesting to investigate whether signatures of a state 
characterized by $f$-electrons in a mixed $f^7$ localized and 1.5 $f$ itinerant 
picture could be observed for berkelium. Such a state would have similarity to the $\delta$-Pu phase, i.e. both
distinct localized features, like deep-lying $f$-states in photoemission, and itinerant characteristics,
like a large density of states at the Fermi level. However, in contrast to $\delta$-Pu the Bk atoms
would possess large magnetic moments, arranged antiferromagnetically.
The present study is restricted to the high-symmetry structures fcc, hcp and dhcp, and thus
it cannot be excluded that around 
the $f$-delocalization transition the energy gain by full delocalization and 
adoption of an appropriate low-symmetry phase outweighs that of only partial delocalization.

The paramagnetic scenario was also investigated for Bk (not shown). 
Energetically, this state is unfavorable (see Table I), but also other aspects of the calculation are in disagreement
with experiment.
The $f^8$ localized shell is found to be the lowest energy state in this case, with an equilibrium volume of
$V=205.1$ a.u., i.e. somewhat larger than the volume of the magnetic phase. However the $f^7$
localized scenario is very close in energy (0.02 eV per atom), 
implying a possible low-pressure ($p\sim 1$ GPa) transition to this state, which furthermore would
be accompanied by a $\sim 10$ \% volume discontinuity. 
Experimentally, there is clear  evidence of large antiferromagnetically ordered
moments.\cite{huray}

\section{Discussion}


\begin{figure}
\begin{center}
\includegraphics[width=90mm,clip]{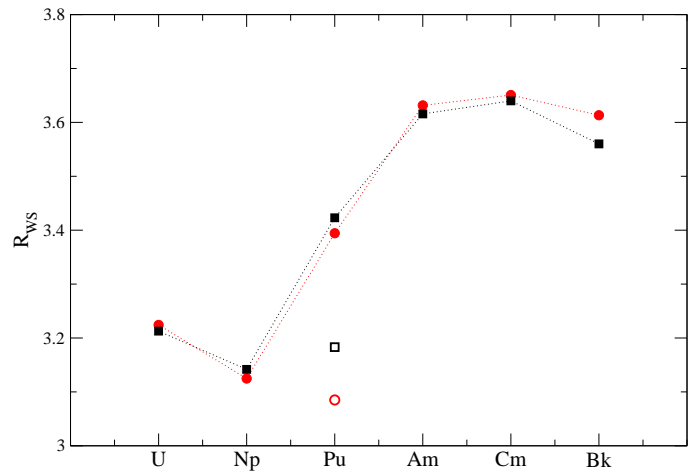}\\
\caption{
\label{Rwstrends}
(Color online)
Trends in the Wigner-Seitz radius through the
actinide series. The (red) circles and (black) squares are calculated (in fcc structure)
and experimental
Wigner-Seitz radii, respectively.
For Pu the $f^4$ non-magnetic ground state
is taken and compared with the experimental Wigner-Seitz radius of 
$\delta$-Pu. The calculated ($f^0$) and experimental $\alpha$-Pu Wigner-Seitz radii are
marked with an open circle and an open square, respectively.
}
\end{center}
\end{figure}
\begin{figure}
\begin{center}
\includegraphics[width=90mm,clip]{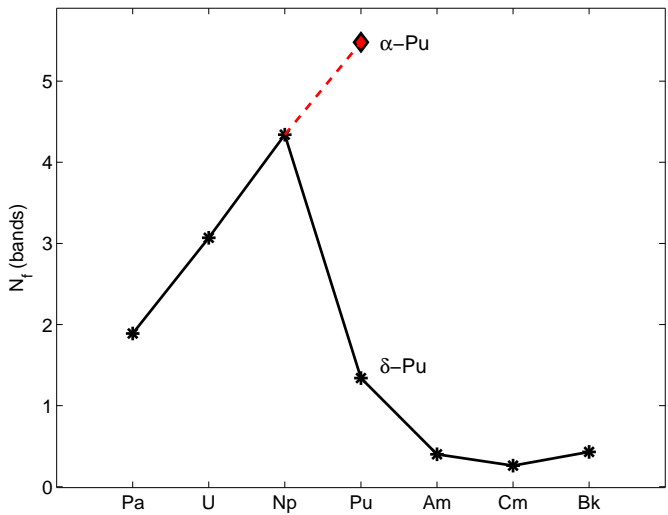}\\
\caption{
\label{Nftrends}
Trends in the the number of band $f$ electrons through the
actinide series. 
For Pu, the $\alpha$- and $\delta$-phases are identified with the $f^0$ and $f^4$ paramagnetic 
states, respectively.
}
\end{center}
\end{figure}

\begin{figure}
\begin{center}
\includegraphics[width=90mm,clip]{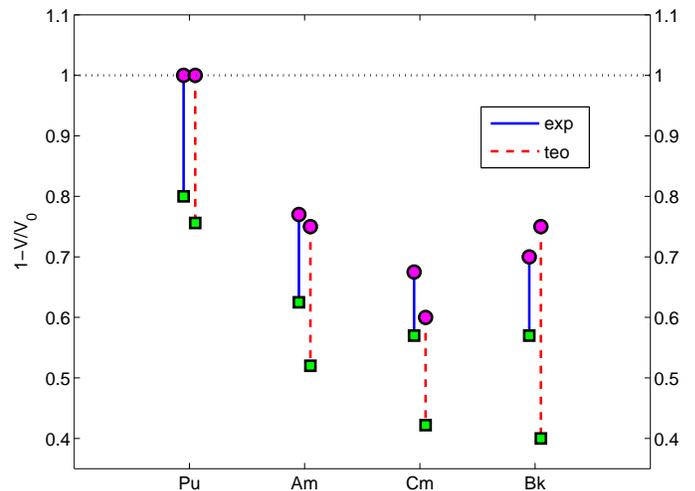}\\
\caption{ 
\label{deloc}
(Color online)
Trends in the ranges of volumes of $f$ electron delocalization in the actinide elements. 
Onset (completion) of delocalization is marked with balls (squares), for each element
with experimental data (blue and full line)
to the left and theoretical data (red and dashed line) to the right.
Volumes are given relative to the (localized) equilibrium volume at ambient conditions. 
 Experimental ranges are defined by the smallest
volumes observed in the high-symmetry (fcc) phase and the largest volumes observed for the
low-symmetry ($\alpha$-U type or similar) phases in high pressure experiments 
(Refs. \onlinecite{heathman}, \onlinecite{landerCm}, \onlinecite{benedict}; 
for Pu the range is defined by the zero pressure
volumes of the $\delta$- and $\alpha$-phases).
The theoretical ranges are defined within the fcc structure only.
}
\end{center}
\end{figure}

Table I and figures \ref{Rwstrends}, \ref{Nftrends}, and \ref{deloc} 
summarize the findings of the present paper.
The rather good agreement between calculated and experimental Wigner-Seitz radii is demonstrated
in Fig. \ref{Rwstrends}. Figure \ref{Nftrends} shows the
calculated number of band-like $f$-electrons in the ground state. 
It is seen to rise steeply from U to $\alpha$-Pu, reflecting the filling of the $f$-band,
before dropping dramatically towards the heavier actinides, reaching its lowest value for
Cm,
after which a small increase is seen again for Bk. For $\delta$-Pu, an intermediate $f$-band filling is
found, which elucidates
 the extraordinary character of
this element, also clear from the essentially vanishing localization/delocalization energy
illustrated in figure \ref{Pujj}.
Correlating Fig. \ref{Rwstrends}
to Fig. \ref{Nftrends}, it follows that the increasing number of band-like $f$ electrons
for U, Np and Pu($f^0$) contributes to the bonding and leads to the 
observed smaller equilibrium volumes,
while larger volumes are obtained for Am, Cm, and Bk,
having a few band-like $f$ electrons.


Several important energy scales may be identified for the actinide elements.
The intraatomic exchange field 
is undoubtedly relevant
but cannot be comprehensively addressed 
by the single-determinant description
underlying the DFT, by which the atomic $S$ and $M_S$ quantum numbers get mixed up. 
In our calculations 
this shortcoming translates
into wrong predictions of magnetic ground state configurations of Pu and Am.
The energy scale for the interatomic exchange interaction is small, favoring the antiferromagnetic arrangement over
the ferromagnetic arrangement with 0.02-0.04 eV 
per atom in Pu, Am, Cm and Bk. 
The localization/delocalization energy scale is larger but depends strongly on volume.\cite{SSC}
As a consequence, $f$-electrons are eventually 
transferred from being localized to
becoming band-like 
when pressure is applied. 
In figure \ref{deloc}, we show the trends in the range of relative volume
over which the $f$-electron delocalization takes place for Pu to
Bk. The volume range where the $f$-electron delocalization occurs is defined in theory by
the lowest volume having the same number of localized electrons as in the ground state
(onset), and the largest volume where a completely delocalized ($f^0$) $f$-manifold 
occurs. In experiment there is no clear-cut definition, but presumably as long as the 
crystal phase remains fcc under compression, the $f$-shell is completely localized, 
whereas a complete delocalization may be assumed to have occurred once the high-pressure
low-symmetry phases of AmIV,\cite{heathman} CmIV,\cite{landerCm} and BkIII\cite{benedict}
are reached. The trends of experiments and theory are in agreement, with Pu at the
delocalization threshold under ambient conditions, and Cm the most stable of the actinide
elements studied. The theoretical volumes of $f$ delocalization completion tend to be smaller
than the experimental ones, which most probably is due to the restriction of the present calculations
to the high-symmetry fcc and hexagonal phases.

A striking result of the calculations presented here is that an excellent account of the
cohesive properties of the localized phases of the actinide elements can only be achieved,
if $\delta$-Pu and Am are described by a paramagnetic phase, 
while a magnetic phase 
is appropriate for Cm and Bk.
The SIC-LSD total energy functional incorrectly favors the formation of a magnetic state
for $\delta$-Pu and Am.
The reason for this shortcoming is probably different for these two elements. In Pu, the $f$-shell is
in a vividly fluctuating state, and presumably any tendency to form interatomic magnetic correlations
is  washed away by quantum fluctuations. 
In Am, the $f^6$ shell is more robust against fluctuations, but here 
the overestimated tendency to form a magnetic state
can be traced back to the inadequacy of the LSD to describe Russell-Saunders
coupling and the formation of Hund's rules ground states, 
as discussed in section III.A. 

Due to its construction, the SIC-LSD can only describe an integer number of localized states
on each atom, and the delocalization process upon compression is modelled as a series of
steps, characterized by one more $f$-electron changing from atomic to band-like behavior.
Such a description can only be considered approximative, and in   a more elaborate theory
the transitions are likely to occur continuously. 
To describe these continuous transitions at finite temperature 
we can construct a mean-field coherent potential approximation
of the fluctuating localization configurations distinguished by different numbers of
localized states. This has already been implemented for cerium,\cite{LSIC}
and a similar study for the actinides is in progress.
Bk may be an exception, due to the
high stability of the $f^7$ shell, and one may envisage a
pressure range where the $f^7$ shell is still localized while the eighth $f$ electron, of
spin minority character, has delocalized. This would be an interesting effect to look for.
A more complete study of the delocalization process would also need to consider the
effects of crystal structure, as it is well known that $f$-electrons once starting to
form bands also prefer low-symmetry structures. This necessitates a full-potential
description of the total energy, which will be subject of future development of the SIC-LSD
method.

Finally, the most important point of Fig.10, namely Pu situated at the
delocalisation threshold under ambient conditions, and trivalent Cm to be the most
stable actinide studied here, is confirmed by a recent DMFT calculation.\cite{shim} 
This SIC-LSD result shows that the SIC approach is capable of
identifying the important valence states across which the DMFT has to fluctuate
dynamically: more or less all possible valence fluctuations for Pu, and the pure
trivalent state for Cm.


\section{Summary}

The cohesive and electronic properties of the actinide elements U, Np, Pu, Am, Cm and Bk have been
investigated with the self-interaction corrected local spin density approximations.
The predicted non-magnetic/magnetic ground states are in agreement with experiment
for the light actinides U, and Np, and the later actinides Cm and Bk, but a magnetic ground state
is wrongly predicted for Am and Pu. 
However, an accurate description has been obtained (see Table I),  
when 
Pu and Am are forced into  a paramagnetic state. 
Cm and Bk are correctly described in an antiferromagnetic arrangement. 
U and Np prefer a non-magnetic delocalized $f$ ground state configuration.
The ground states at zero temperature and pressure are
the trivalent Am($f^6$), Cm($f^7$), and Bk($f^8$), while Pu shows almost degeneracy between 
any of the localization configurations from $f^0$ to $f^4$. The latter signals a true ground state
of a much more complex nature than can be described with the present approach, characterized by
quantum fluctuations between any of these states. 
If the stabilization of Pu($f^4$) with high
temperature or Ga alloying is envisaged, this state may be identified with the $\delta$-phase of
Pu, with an intermediate number of localized and itinerant $f$-electrons and an accompanying
 large density of states at the Fermi level.
The delocalization of the $f$-shell under compression has been studied, and the delocalization
has been found to occur over a range of volumes, with the established trends in good 
agreement with experiments. 
However, to describe the actual high pressure crystal sructures and the
structural transitions, a more accurate implementation of SIC-LSD, most
notably including full non-spherical potentials, is necessary.

\section{acknowledgements}

Discussion with B. Johansson are greatly acknowledged.
This work was partially funded by the EU Research Training Network
(contract:HPRN-CT-2002-00295) 'Ab-initio Computation of Electronic 
Properties of $f$-electron Materials'. AS acknowledges support from the Danish Center for Scientific
Computing.  Work of LP was
sponsored by the Office of Basic Energy Sciences, U.S. Department of Energy.
A portion of this research was conducted at the Center for Nanophase Materials
Sciences, which is sponsored at Oak Ridge National Laboratory by the Division
of Scientific User Facilities, U.S. Department of Energy.

\end{document}